\documentclass[10pt,a4paper]{article}
\usepackage{amssymb,amsfonts}

\begin{document}

\title{An Algebraic Formulation of Level
One Wess-Zumino-Witten Models}
\author{Jens B{\"o}ckenhauer\\II. Institut f{\"u}r
Theoretische Physik, Universit{\"a}t Hamburg\\
Luruper Chaussee 149, D-22761 Hamburg}
\maketitle

\begin{abstract}
The highest weight modules of the chiral algebra of
orthogonal WZW models at level one possess a realization
in fermionic representation spaces; the Kac-Moody
and Virasoro generators are represented as unbounded
limits of even CAR algebras. It is shown that the
representation theory of the underlying even CAR
algebras reproduces precisely the sectors of the
chiral algebra. This fact allows to develop a
theory of local von Neumann algebras on the
punctured circle, fitting nicely in the
Doplicher-Haag-Roberts framework. The relevant
localized endomorphisms which generate the
charged sectors are explicitly constructed by
means of Bogoliubov transformations. Using CAR
theory, the fusion rules in terms of sector
equivalence classes are proven.
\end{abstract}

\newcommand{\I}{{\rm i}}
\newcommand{\E}{{\rm e}}
\newcommand{\SO}{{\widehat{\rm\bf so}(N)}}
\newtheorem{definition}{Definition}[section]
\newtheorem{lemma}[definition]{Lemma}
\newtheorem{corollary}[definition]{Corollary}
\newtheorem{theorem}[definition]{Theorem}
\newtheorem{proposition}[definition]{Proposition}

\section{Introduction}
In local quantum field theory one considers a Hilbert
space ${\cal H}$ of physical states which decomposes
into orthogonal subspaces ${\cal H}_J$ (superselection
sectors) so that observables do not make transitions
between the sectors. The subspaces ${\cal H}_J$ carry
inequivalent, irreducible representations of the
observable algebra $\mathfrak{A}_{\rm loc}$, possibly with
some multiplicities (see \cite{Haag} for an overview).
Among the superselection sectors,
there is a distinguished sector ${\cal H}_0$ which
contains the vacuum vector $|\Omega_0\rangle$ and
carries the vacuum representation $\pi_0$.

The starting point in the algebraic approach to quantum
field theory \cite{DHR1,DHR2,Haag} is the quasilocal
observable algebra $\mathfrak{A}_{\rm loc}$ which
is usually defined as the $C^*$-inductive limit of
the net of local von Neumann algebras
$\{{\cal R}({\cal O}),{\cal O}\in\mathfrak{K}\}$,
where $\mathfrak{K}$ denotes the set of open double cones
in $D$ dimensional Minkowski space. The
Doplicher-Haag-Roberts (DHR) criterion selects only
those representations $\pi_J$ which become equivalent
to the vacuum representation in restriction to the
algebra $\mathfrak{A}({\cal O}')$ of the causal complement
${\cal O}'$ for some sufficiently large double cone
${\cal O}$. ($\mathfrak{A}({\cal O}')$ is the
norm-closure of the union of all ${\cal R}({\cal O}_1)$,
${\cal O}_1\subset{\cal O}'$.) The DHR selection
criterion leads to the description of sectors
(unitary equivalence classes $[\pi_J]$)  by
localized endomorphisms $\varrho_J$ of
$\mathfrak{A}_{\rm loc}$, $\varrho_J(A)=A$ for all
$A\in\mathfrak{A}({\cal O}')$, so that all physical
information is contained in the vacuum sector:
$\pi_J\simeq\pi_0\circ\varrho_J$. This leads to
the important fact that DHR sectors can be composed,
i.e.~one has a product of sectors
$[\pi_J]\times[\pi_{J'}]=[\pi_0\circ\varrho_J\varrho_{J'}]$;
so one can derive fusion rules, given by the composition
of localized endomorphisms.

While being an abstract and mathematically rigorous
setting, the DHR analysis appeared to be difficult
to be applied to concrete quantum field theory models.
In the last years two-dimensional  conformal quantum
field theory (CQFT) turned out to be a hopeful
area of application. Many of the features of CQFT
appear to be closely related to the abstract structures
of the algebraic setting \cite{Fuchs1}. For instance,
the chiral algebra can be considered as observable
algebra and its highest weight modules play the role
of superselection sectors. Then the conformal fusion
rules seem to be the natural counterpart of the
product of sectors realized by the composition of
localized endomorphisms in the algebraic approach.
However, there is no mathematically precise
prescription known how to translate the objects
of CQFT into the framework of algebraic quantum
field theory. Indeed, the attempts to incorporate
CQFT models in this framework sometimes seemed to
require some deviations of the DHR program;
operator algebras of observables and endomorphisms
were constructed in \cite{MS1,FGV},
however, the use of non-localized endomorphisms,
non-faithful representations and algebras containing
unbounded elements violated the canonical DHR
framework. As a consequence, the composition of
non-localized endomorphisms could not be generalized
to the fusion of sectors i.e.~of equivalence classes
$[\pi_J]$ of representations.

Inspired by a paper of Fuchs, Ganchev and
Vecserny\'{e}s \cite{FGV}, we give a formulation
close to the DHR program for a special class of
conformal models, the
${{\rm\bf so}(N)}$-Wess-Zumino-Witten
(WZW) models at level one.
Because the Kac-Moody and Virasoro generators of
their chiral algebra can be built as infinite series
of fermion bilinears in representation spaces of the
canonical anticommutation relations (CAR) one expects
that the sectors of the chiral algebra (highest
weight modules) find their analogue in irreducible
representations of the underlying even CAR algebras.
Using results of Araki's selfdual CAR algebra, we
show that there is indeed a one-to-one correspondence.
So we can restrict our attention to algebras of
bounded operators and get a theory of local
$C^*$-algebras ${\cal A}(I)$ on the circle
($I\subset S^1$). We then construct localized
endomorphisms by means of Bogoliubov transformations
and show that we can extend our representations and
endomorphisms to a net of local von Neumann algebras
${\cal R}(I)$ which generate a quasilocal algebra
$\mathfrak{A}_{\rm loc}$ on the punctured circle.
Thus, up to the replacement of double cones by
intervals, we do not leave the DHR framework.
Hence now one can deduce fusion rules in terms of
sector equivalence classes by computing the
composition of special representative localized
endomorphisms. Using our examples, we indeed
rediscover the well-known WZW fusion rules.

\section{The Chiral Algebra of Level 1 WZW Models}
As already mentioned, in two-dimensional CQFT
the analog role of the observable
algebra is played by the chiral algebra
which is the chiral half of the symmetry algebra.
In WZW theory we consider here,
the chiral algebra ${\cal L}$ is the semi-direct
product of an untwisted Kac-Moody algebra $\SO_k$,
generated by ``currents'' $J_m^\alpha$,
$\alpha=1,2,\ldots,\frac{1}{2}N(N-1)$, $m\in{\mathbb Z}$,
$N\in{\mathbb N}$ fixed, and the Virasoro
algebra ${\sf Vir}_c$, generated by
$L_n$, $n\in{\mathbb Z}$; in our models the level and
central charge are fixed to be $k=1$ and $c=N/2$,
\begin{equation}
\label{L}
{\cal L} = \SO_1 \rtimes {\sf Vir}_{N/2}.
\end{equation}
Now let $T^\alpha$, $\alpha=1,2,\ldots,\frac{1}{2}N(N-1)$
be real antisymmetric matrix generators of the
finite-dimensional Lie algebra ${{\rm\bf so}(N)}$
in the defining (vector) representation. We denote by
$f_\gamma^{\alpha\beta}$ the structure constants and
by $\kappa^{\alpha\beta}$ the Cartan-Killing form
of ${{\rm\bf so}(N)}$, i.e.~$[T^\alpha,T^\beta]=
f_\gamma^{\alpha\beta}T^\gamma$ (summation over
$\gamma$) and
$\kappa^{\alpha\beta}={\rm tr}(T^\alpha T^\beta)$.
Then the commutation relations read
\begin{eqnarray}
\label{CC1}
[J_m^\alpha,J_n^\beta] &=& f_\gamma^{\alpha\beta}
J_{m+n}^\gamma + m\kappa^{\alpha\beta} \delta_{m,-n},\\
\label{CC2}
[L_m,L_n] &=& (m-n)L_{m+n} + \frac{N}{24}(m^3-m)
\delta_{m,-n},\\
\label{CC3}
[L_m,J_n^\alpha] &=& -n J_{m+n}^\alpha.
\end{eqnarray}
The representation theory of ${\cal L}$ is well
known \cite{GO,Kac}, for $N$ even,
i.e.~$N\in 2{\mathbb N}$, there are four different
integrable highest weight modules, the basic
(denoted by $0$ or $1$), the vector (v) and two
different spinor modules (s and c); for $N$ odd,
i.e.~$N\in 2{\mathbb N}_0+1$ there is only the basic
($0,1$), the vector (v) and one spinor ($\sigma$)
module. (The case $N=1$ reproduces formally the
Ising model.) These modules correspond to
$L_0$-eigenvalues $h_0=0$, $h_{\rm v}=1/2$ and
$h_{\rm s}=h_{\rm c}=h_\sigma=N/16$ of their highest
weight vectors. We will realize
these modules in representation spaces
${\cal H}_{\rm NS}$ and ${\cal H}_{\rm R}$ of CAR. The
${\sf Vir}_{N/2}$ and $\SO_1$ Kac-Moody generators
then become expressions as infinite series of
normal ordered fermion bilinears, that means they are
represented as unbounded limits of even CAR algebras.
For this purpose we introduce $N$ Majorana fields
$\psi^i$, $i=1,2,\ldots,N$, on the circle $S^1$,
with hermiticity condition
\[ (\psi^i(z))^* =  \psi^i(z) \]
and satisfying anticommutation relations
\[ \{ \psi^i(z),\psi^j(z') \}
= 2\pi \I z \,\delta^{i,j}\delta(z-z'). \]
Consider an $N$-component $L^2$-function on the circle,
$f=(f^i)_{i=1,2,\ldots,N}\in{\cal K}=
L^2(S^1;{\mathbb C}^N)$. Define an antiunitary
involution $\Gamma$ by component-wise complex conjugation,
$\Gamma f=(\overline{f^i})_{i=1,2,\ldots,N}$.
Then smeared objects
\[ B(f) = \sum_{j=1}^N \oint_{S^1}
\frac{{\rm d}z}{2\pi \I z} f^j(z) \psi^j(z) \]
obey the defining relations of the canonical generators
of Araki's \cite{Ara1,Ara2} selfdual CAR algebra
${\cal C}({\cal K},\Gamma)$,
\[ \{ B(f)^*,B(g) \} = \langle f,g \rangle {\bf 1},
\qquad B(f)^* = B(\Gamma f), \qquad f,g\in{\cal K}, \]
with the canonical scalar product
$\langle\cdot,\cdot\rangle$ on ${\cal K}$,
\[ \langle f,g \rangle = \sum_{j=1}^N \oint_{S^1}
\frac{{\rm d}z}{2\pi \I z}
\overline{f^j(z)} g^j(z). \]
The selfdual CAR algebra is discussed in the
following section.

\section{The Selfdual CAR Algebra}
Let ${\cal K}$ be some Hilbert space with an
antiunitary involution $\Gamma$ (complex conjugation),
$\Gamma^2={\bf 1}$, which obeys
\[ \langle \Gamma f, \Gamma g \rangle = \langle g ,
f \rangle, \qquad f,g\in{\cal K}. \]
The selfdual CAR algebra ${\cal C}({\cal K},\Gamma)$
is defined to be the $C^*$-norm closure of the algebra
which is generated by the range
of a linear mapping $B:f\mapsto B(f)$,
such that
\[ \{ B(f)^*,B(g) \} = \langle f,g \rangle {\bf 1},
\qquad B(f)^*=B(\Gamma f), \qquad f,g\in{\cal K}. \]
holds. The $C^*$-norm satisfies
\begin{equation}
\label{Cnorm}
\|B(f)\| \le \|f\|, \qquad f \in {\cal K}.
\end{equation}
The states of ${\cal C}({\cal K},\Gamma)$ we are
interested in are called quasifree states. By definition,
a quasifree state $\omega$ fulfills for $n\in {\mathbb N}$
\begin{eqnarray*}
\omega (B(f_1) \cdots B(f_{2n+1})) &=& 0, \\
\omega (B(f_1) \cdots B(f_{2n})) &=&
(-1)^\frac{n(n-1)}{2} \sum_\sigma
\mbox{\rm sign} \sigma\prod_{j=1}^n \omega
(B(f_{\sigma(j)})B(f_{\sigma(n+j)}))\,\,\,
\end{eqnarray*}
where the sum runs over all permutations
$\sigma \in {\cal S}_{2n}$ with the property
\[ \sigma (1) < \sigma (2) < \cdots < \sigma (n),
\qquad \sigma (j) < \sigma (j+n), \qquad j=1,\ldots,n.\]
Clearly, quasifree states are completely characterized by
their two point functions. Moreover, there is a
one-to-one correspondence between the set of quasi\-free
states and the set
\[ {\cal Q}({\cal K},\Gamma)=\{S\in\mathfrak{B}({\cal K})
\,|\, S=S^*,\, 0 \le S \le {\bf 1},\,
S + \overline{S} = {\bf 1} \}, \]
(we have used the notation $\overline{A}=\Gamma A\Gamma$
for bounded operators $A\in\mathfrak{B}({\cal K})$)
given by the formula
\begin{equation}
\label{phi-S}
\omega (B(f)^* B(g)) = \langle f,Sg \rangle .
\end{equation}
So it is convenient to denote the quasifree state
characterized by Eq.~(\ref{phi-S}) by $\omega_S$.
The projections in ${\cal Q}({\cal K},\Gamma)$
are called basis projections or polarizations.
For a basis projection $P$, the state $\omega_P$
is pure and is called a Fock state. The corresponding
GNS representation $({\cal H}_P,\pi_P,|\Omega_P\rangle)$
is irreducible, it is called the Fock representation.
The space ${\cal H}_P$ can be canonically identified
with the antisymmetric Fock space
${\cal F}_-(P{\cal K})$.
There is an important quasiequivalence criterion
for GNS representations of quasifree states.
Quasiequivalence will be denoted by "$\approx$"
and unitary equivalence by "$\simeq$". Let us denote
by $\mathfrak{J}_2({\cal K})$ the ideal of Hilbert-Schmidt
operators in $\mathfrak{B}({\cal K})$ and for
$A\in\mathfrak{B}({\cal K})$ by $[A]_2$ its
Hilbert-Schmidt equivalence class
$[A]_2=A+\mathfrak{J}_2({\cal K})$.
Araki proved \cite{Ara1,Ara2}
\begin{theorem}
\label{Araki}
For quasifree states $\omega_{S_1}$ and
$\omega_{S_2}$ of ${\cal C}({\cal K},\Gamma)$
we have quasiequiv\-alence $\pi_{S_1}\approx\pi_{S_2}$
if and only if
$[S_1^\frac{1}{2}]_2=[S_2^\frac{1}{2}]_2$.
\end{theorem}
Next we define the set
\[ {\cal I}({\cal K},\Gamma) = \{ V\in \mathfrak{B}({\cal K})
\,|\, V^*V={\bf 1},\,\, V=\overline{V} \} \]
of Bogoliubov operators. Bogoliubov operators
$V\in {\cal I}({\cal K},\Gamma)$ induce unital
$\ast$-endomorphisms $\varrho_V$ of
${\cal C}({\cal K},\Gamma)$, defined by their action
on the canonical generators,
\[ \varrho_V(B(f))=B(Vf). \]
Moreover, if $V\in {\cal I}({\cal K},\Gamma)$ is surjective
(i.e.~unitary), then $\varrho_V$ is an automorphism.
A quasifree state, composed with a
Bogoliubov endomorphism is again a quasifree state,
namely we have $\omega_S\circ\varrho_V=\omega_{V^*SV}$.
In the following we are interested in representations
of the form $\pi_P\circ\varrho_V$ instead of GNS
representations $\pi_{V^*PV}$ of states
$\omega_{V^*PV}=\omega_P\circ\varrho_V$.
Indeed, the former are multiples of the latter,
in particular, we have \cite{Binnenneu,Rideau}
\begin{equation}
\label{cyclics}
\pi_P\circ\varrho_V\simeq 2^{N_V}\pi_{V^*PV},
\qquad N_V={\rm dim}({\rm ker}V^*\cap P{\cal K}).
\end{equation}
Thus, the identification of the Hilbert-Schmidt
equivalence class $[(V^*PV)^\frac{1}{2}]_2$ is
the identification of the quasiequivalence class
of $\pi_P\circ\varrho_V$. For the identification
of the unitary equivalence class, we need a
decomposition of $\pi_P\circ\varrho_V$ into
irreducible subrepresentations which will now be
elaborated. A projection $E\in\mathfrak{B}({\cal K})$
with the property that $E\overline{E}=0$ and that
ker$(E+\overline{E})=\mathbb{C}e_0$ with a
$\Gamma$-invariant unit vector $e_0\in{\cal K}$
is called a partial basis projection with
$\Gamma$-codimension 1. Note that $E$ defines a
Fock representation
$({\cal H}_E,\pi_E,|\Omega_E\rangle)$ of
${\cal C}((E+\overline{E}){\cal K},\Gamma)$.
Following Araki, pseudo Fock representations
$\pi_{E,+}$ and $\pi_{E,-}$ of
${\cal C}({\cal K},\Gamma)$ are defined in
${\cal H}_E$ by
\begin{equation}
\label{piE}
\pi_{E,\pm}(B(f)) = \pm \frac{1}{\sqrt{2}}
\langle e_0,f \rangle Q_E(-1) +
\pi_E(B((E+\overline{E})f), \qquad f\in{\cal K},
\end{equation}
where $Q_E(-1)\in\mathfrak{B}({\cal K})$ is the
unitary, self-adjoint implementer of the automorphism
$\alpha_{-1}$ of ${\cal C}({\cal K},\Gamma)$
defined by $\alpha_{-1}(B(f))=-B(f)$ (which
restricts also to an automorphism of
${\cal C}((E+\overline{E}){\cal K},\Gamma)$).
Pseudo Fock representations $\pi_{E,+}$ and
$\pi_{E,-}$ are inequivalent and irreducible.
Araki proved
\cite{Ara1}
\begin{lemma}
\label{pFock}
Let $E$ be a partial basis projection with
$\Gamma$-codimension 1, and let $e_0\in{\cal K}$
be a $\Gamma$-invariant unit vector of
${\rm ker}(E+\overline{E})$. Define
$S\in{\cal Q}({\cal K},\Gamma)$ by
\begin{equation}
\label{mittel}
S = \frac{1}{2} |e_0\rangle\langle e_0| + E.
\end{equation}
Then a GNS representation
$({\cal H}_S,\pi_S,|\Omega_S\rangle)$ of the
quasifree state $\omega_S$ is given by the
direct sum of two inequivalent, irreducible
pseudo Fock representations,
\begin{equation}
({\cal H}_S,\pi_S,|\Omega_S\rangle) =
\left( {\cal H}_E \oplus {\cal H}_E,\pi_{E,+}
\oplus \pi_{E,-}, \frac{1}{\sqrt{2}} (|\Omega_E
\rangle \oplus |\Omega_E \rangle) \right) .
\end{equation}
\end{lemma}
It was the observation in \cite{Decom} (see also
\cite{Binnenneu}) that only Fock and pseudo
Fock representations appear in the decomposition of
representations $\pi_P\circ\varrho_V$ if the
Bogoliubov operator has finite corank.
\begin{theorem}
\label{evenodd}
Let $P$ be a basis projection and
let $V$ be a Bogoliubov operator with
$M_V=\mbox{\rm dim ker}V^* < \infty$. If $M_V$ is
an even integer we have (with notations as above)
\begin{equation}
\pi_P\circ\varrho_V \simeq 2^\frac{M_V}{2} \pi_{P'}
\end{equation}
where $\pi_{P'}$ is an (irreducible) Fock representation.
If $M_V$ is odd then we have
\begin{equation}
\label{decomodd}
\pi_P\circ\varrho_V \simeq 2^\frac{M_V-1}{2}
(\pi_{E,+} \oplus \pi_{E,-})
\end{equation}
where $\pi_{E,+}$ and $\pi_{E,-}$ are inequivalent
(irreducible) pseudo Fock representations.
\end{theorem}
We define the even algebra ${\cal C}({\cal K},\Gamma)^+$
to be the subalgebra of $\alpha_{-1}$-fixpoints,
\[ {\cal C}({\cal K},\Gamma)^+ = \{ x \in {\cal C}
({\cal K},\Gamma) \,\,|\,\, \alpha_{-1}(x)=x \}. \]
We now are interested in what happens when our
representations of ${\cal C}({\cal K},\Gamma)$
are restricted to the even algebra. For basis
projections $P_1,P_2$, with $[P_1]_2=[P_2]_2$,
Araki and D.E.~Evans \cite{AE} defined an index,
taking values $\pm 1$,
\[ \mbox{ind}(P_1,P_2)= (-1)^{{\rm dim}(
P_1{\cal K}\cap({\bf 1}-P_2){\cal K})}. \]
The automorphism $\alpha_{-1}$ leaves any
quasifree state $\omega_S$ invariant. Hence
$\alpha_{-1}$ is implemented in $\pi_S$. In particular,
in a Fock representation $\pi_P$, $\alpha_{-1}$
extends to an automorphism $\bar{\alpha}_{-1}$ of
$\pi_P({\cal C}({\cal K},\Gamma))''=
\mathfrak{B}({\cal H}_P)$. The following proposition
is taken from \cite{Ara2}.
\begin{proposition}
\label{ind}
Let $U\in{\cal I}({\cal K},\Gamma)$ be a unitary
Bogoliubov operator and let $P$ be a basis
projection such that $[P]_2=[U^*PU]_2$.
Denote by $Q_P(U)\in\mathfrak{B}({\cal H}_P)$ the
unitary which implements $\varrho_U$ in $\pi_P$. Then
\begin{equation}
\bar{\alpha}_{-1}(Q_P(U)) = \sigma(U) Q_P(U),\qquad
\sigma(U)=\pm 1.
\end{equation}
In particular, $\sigma(U)=\mbox{\rm ind}(P,U^*PU)$.
Moreover, given two unitaries
$U_1,U_2\in{\cal I}({\cal K},\Gamma)$ of this type,
$\sigma$ is multiplicative,
$\sigma(U_1U_2)=\sigma(U_1)\sigma(U_2)$.
\end{proposition}
Furthermore, one has \cite{AE,Ara2}
\begin{theorem}
\label{resteven}
Restricted to the even algebra
${\cal C}({\cal K},\Gamma)^+$, a Fock representation
$\pi_P$ splits into two mutually inequivalent, irreducible
subrepresentations,
\begin{equation}
\pi_P|_{{\cal C}({\cal K},\Gamma)^+}=\pi_P^+\oplus
\pi_P^-.
\end{equation}
Given two basis projections $P_1,P_2$, then
$\pi_{P_1}^\pm \simeq \pi_{P_2}^\pm$
if and only if $[P_1]_2=[P_2]_2$ and
{\rm ind}$(P_1,P_2)=+1$, and
$\pi_{P_1}^\pm \simeq \pi_{P_2}^\mp$
if and only if $[P_1]_2=[P_2]_2$ and
{\rm ind}$(P_1,P_2)=-1$.
\end{theorem}
For some real $v\in{\cal K}$, i.e.~$\Gamma v=v$, and
$\|v\|=1$ define $U\in{\cal I}({\cal K},\Gamma)$ by
\begin{equation}
\label{Uvec}
U = 2|v\rangle\langle v|-{\bf 1}.
\end{equation}
Then $\varrho_U$ is implemented in each Fock
representation $\pi_P$ by the unitary self-adjoint
$Q_P(U)=\sqrt{2}\pi_P(B(v))$, since $\varrho_U$ is
implemented in ${\cal C}({\cal K},\Gamma)$ by
$q(U)=\sqrt{2}B(v)$,
\begin{eqnarray*}
q(U)B(f)q(U)
&=& 2B(v)B(f)B(v) \\
&=& 2\{B(v),B(f)\}B(v) - 2B(f)B(v)B(v) \\
&=& 2\langle v,f \rangle B(v)  - B(f) \\
&=& B( 2\langle v, f \rangle v - f)  \\
&=& B(Uf).
\end{eqnarray*}
Hence $\sigma(U)=-1$ and we immediately have the
following
\begin{corollary}
\label{Bogvec}
Let $U\in{\cal I}({\cal K},\Gamma)$ be as in
Eq.~(\ref{Uvec}), then for each Fock
representation $\pi_P$ in restriction to
${\cal C}({\cal K},\Gamma)^+$ we have equivalence
$\pi_P^\pm \circ \varrho_U \simeq \pi_P^\mp$.
\end{corollary}
It was proven in \cite{Decom} that pseudo Fock
representations $\pi_{E,+}$ and $\pi_{E,-}$ of
Theorem \ref{evenodd}, Eq.~(\ref{decomodd}),
when restricted to the even algebra, remain
irreducible but become equivalent.
Summarizing we obtain
\begin{theorem}
\label{restodd}
With notations of Theorem \ref{evenodd}, a representation
$\pi_P\circ\varrho_V$ restricts as follows to the even
algebra ${\cal C}({\cal K},\Gamma)^+$: If $M_V$ is even
we have
\begin{equation}
\pi_P\circ\varrho_V|_{{\cal C}({\cal K},\Gamma)^+}
\simeq 2^\frac{M_V}{2}(\pi_{P'}^+\oplus\pi_{P'}^-)
\end{equation}
with $\pi_{P'}^+$ and $\pi_{P'}^-$ mutually
inequivalent and irreducible. If $M_V$ is odd, then
\begin{equation}
\pi_P\circ\varrho_V|_{{\cal C}({\cal K},\Gamma)^+}
\simeq 2^\frac{M_V+1}{2} \pi
\end{equation}
with $\pi$ irreducible.
\end{theorem}

\section{Construction of the Representation Spaces}
Now we are ready to build our ${\cal L}$-modules
as representation spaces of
${\cal C}({\cal K},\Gamma)$. Recall that
${\cal K}=L^2(S^1;{\mathbb C}^N)\equiv
L^2(S^1)\otimes{\mathbb C}^N$ in our model.
So we obtain two (Fourier) orthonormal bases (ONB)
\[ \left\{e_r^i,r\in{\mathbb Z}+\frac{1}{2},
i=1,2,\ldots,N \right\},\qquad
\{e_n^i,n\in{\mathbb Z},i=1,2,\ldots,N\} \]
by the definition
\[ e_p^i = e_p \otimes u^i,\qquad
p\in\frac{1}{2}{\mathbb Z},\qquad i=1,2,\ldots,N, \]
where $e_p\in L^2(S^1)$ are defined by
$e_p(z)=z^p$ and $u^i$ denote the canonical unit
vectors of ${\mathbb C}^N$. Consider
$P_{\rm NS},S_{\rm R}\in{\cal Q}({\cal K},\Gamma)$, the
Neveu-Schwarz operator
\[ P_{\rm NS}= \sum_{i=1}^N
\sum_{r\in {\mathbb N}_0+\frac{1}{2}}
|e_{-r}^i \rangle \langle e_{-r}^i | \]
is a basis projection, the Ramond operator
\[ S_{\rm R}= \sum_{i=1}^N \left( \frac{1}{2}
|e_0^i\rangle\langle e_0^i| + \sum_{n\in {\mathbb N}}
|e_{-n}^i \rangle\langle e_{-n}^i| \right) \]
is not. Let us denote by
$({\cal H}_{\rm NS},\pi_{\rm NS},|\Omega_{\rm NS}\rangle)$
and
$({\cal H}_{\rm R},\pi_{\rm R},|\Omega_{\rm R}\rangle)$ the GNS
representations of the associated quasifree states
$\omega_{P_{\rm NS}}$ and $\omega_{S_{\rm R}}$, respectively,
i.e.~in order to avoid double indices, we write
$\pi_{\rm NS}$ instead of $\pi_{P_{\rm NS}}$, $\pi_{\rm R}$
instead of $\pi_{S_{\rm R}}$ etc.
Further define for $i=1,2,\ldots,N$ the Fourier modes
\[ b_r^i=\pi_{\rm NS}(B(e_r^i)),\qquad r\in{\mathbb Z}
+\frac{1}{2};\qquad b_n^i=\pi_{\rm R}(B(e_n^i)),
\qquad n\in{\mathbb Z}, \]
such that we have CAR
$\{b_r^i,b_s^j\}=\delta^{i,j}\delta_{r,-s}{\bf 1}$
in ${\cal H}_{\rm NS}$ and
$\{b_m^i,b_n^j\}=\delta^{i,j}\delta_{m,-n}{\bf 1}$
in ${\cal H}_{\rm R}$. It follows that
\[ b_r^i|\Omega_{\rm NS}\rangle = 0, \quad r>0,\qquad
b_n^i |\Omega_{\rm R} \rangle = 0, \quad n>0.\]
Finite particle vectors
\begin{equation}
\label{V-NS}
b_{-r_m}^{i_m} \cdots b_{-r_2}^{i_2}b_{-r_1}^{i_1}
|\Omega_{\rm NS}\rangle, \qquad r_l\in{\mathbb N}_0+\frac{1}{2},
\qquad i_l=1,2,\ldots,N,
\end{equation}
and
\begin{equation}
\label{V-R}
b_{-n_m}^{i_m} \cdots b_{-n_2}^{i_2}b_{-n_1}^{i_1}
|\Omega_{\rm R}\rangle, \qquad n_l\in{\mathbb N}_0,
\qquad i_l=1,2,\ldots,N,
\end{equation}
are total in ${\cal H}_{\rm NS}$ and ${\cal H}_{\rm R}$
i.e. finite linear combinations produce dense subspaces
${\cal H}_{\rm NS}^{\rm fin}$ and
${\cal H}_{\rm R}^{\rm fin}$, respectively.
Denoting normal ordering by colons,
\[ :b_p^i b_q^j:\,\,\, =
\left\{ \begin{array}{rl}
b_p^i b_q^j & p<0 \\ -b_q^j b_p^i & p\ge 0
\end{array} \right. , \qquad p,q\in
\frac{1}{2} {\mathbb Z}, \qquad i,j=1,2,\ldots,N, \]
an action of ${\cal L}$ in
${\cal H}_{\rm NS}^{\rm fin}$ is defined
by the formul{\ae}
\begin{eqnarray*}
J_m^\alpha \longmapsto J_m^{{\rm NS},\alpha} &=&
\frac{1}{2} \sum_{i,j=1}^N (T^\alpha)_{i,j}
\sum_{r\in{\mathbb Z}+\frac{1}{2}}
:b_r^i b_{m-r}^j:,\\
L_m \longmapsto L_m^{\rm NS} &=& -\frac{1}{2}
\sum_{i=1}^N \sum_{r\in{\mathbb Z}+\frac{1}{2}}
\left( r-\frac{m}{2} \right)
:b_r^i b_{m-r}^i:,
\end{eqnarray*}
and in ${\cal H}_{\rm R}^{\rm fin}$ by
\begin{eqnarray*}
J_m^\alpha \longmapsto J_m^{{\rm R},\alpha} &=&
\frac{1}{2} \sum_{i,j=1}^N (T^\alpha)_{i,j}
\sum_{n\in{\mathbb Z}}
:b_n^i b_{m-n}^j:,\\
L_m \longmapsto L_m^{\rm R} &=& -\frac{1}{2}
\sum_{i=1}^N \sum_{n\in{\mathbb Z}}
\left( n-\frac{m}{2} \right)
:b_n^i b_{m-n}^i:
+ \frac{N}{16} \delta_{m,0}.
\end{eqnarray*}
Clearly, these infinite series do not converge in
norm, however, using CAR of the $b_p^i$, they reduce
to finite sums when acting on finite particle vectors
(\ref{V-NS}) and (\ref{V-R}). Thus
$J_m^{{\rm NS},\alpha},L_n^{\rm NS}$ and
$J_m^{{\rm R},\alpha},L_n^{\rm R}$
are indeed well-defined on ${\cal H}_{\rm NS}^{\rm fin}$
and ${\cal H}_{\rm R}^{\rm fin}$, respectively. Relations
(\ref{CC1}), (\ref{CC2}) and (\ref{CC3}) follow
also by direct computation. In ${\cal H}_{\rm NS}$ we
have states
$|0\rangle_{\rm NS}=|\Omega_{\rm NS}\rangle$ and
$|i\rangle_{\rm NS}=
b_{-\frac{1}{2}}^i|\Omega_{\rm NS}\rangle$,
$i=1,2,\ldots,N$, which are eigenvectors of $L_0^{\rm NS}$,
\[ L_0^{\rm NS}|0\rangle_{\rm NS} = 0, \qquad L_0^{\rm NS}
|i\rangle_{\rm NS} = \frac{1}{2} |i\rangle_{\rm NS}. \]
In ${\cal H}_{\rm R}$ we have $2^N$ independent states
$|0\rangle_{\rm R}=|\Omega_{\rm R}\rangle$ and
\[ |i_l,\ldots,i_2,i_1\rangle_{\rm R} =
b_0^{i_l} \cdots b_0^{i_2} b_0^{i_1}
|\Omega_{\rm R}\rangle,
\qquad 1 \le i_1 < i_2 < \cdots < i_l \le N, \]
satisfying
\[ L_0^{\rm R} |0\rangle_{\rm R} =
\frac{N}{16} |0\rangle_{\rm R},
\qquad L_0^{\rm R} |i_l,\ldots,i_2,i_1\rangle_{\rm R}
= \frac{N}{16} |i_l,\ldots,i_2,i_1\rangle_{\rm R}. \]
As ${\cal L}$-modules ${\cal H}_{\rm NS}^{\rm fin}$ and
${\cal H}_{\rm R}^{\rm fin}$ are not irreducible. It is
known that ${\cal H}_{\rm NS}^{\rm fin}$
splits up into the direct sum of the basic
and the vector module, while
${\cal H}_{\rm R}^{\rm fin}$ decomposes into
the direct sum of $2^\frac{N}{2}$ spinor (s) and
$2^\frac{N}{2}$ conjugate spinor (c) modules
if $N$ is even and into $2^\frac{N+1}{2}$ spinor
modules ($\sigma$) if $N$ is odd. Using our previous
results of CAR theory, we can easily verify that
exactly the same happens if we restrict the
representations $\pi_{\rm NS}$ ($\pi_{\rm R}$) of
${\cal C}({\cal K},\Gamma)$ in ${\cal H}_{\rm NS}$
(${\cal H}_{\rm R}$) to the even subalgebra
${\cal C}({\cal K},\Gamma)^+$: Since $P_{\rm NS}$
is a basis projection we have by Theorem
\ref{resteven}
\begin{equation}
\label{NS+}
\pi_{\rm NS}|_{{\cal C}({\cal K},\Gamma)^+} =
\pi_{\rm NS}^+ \oplus \pi_{\rm NS}^-.
\end{equation}
Now $\pi_{\rm NS}^+$ acts in the even Fock space
\cite{Ara2} which corresponds to the basic
module. Thus we may use the same symbols which
label the sectors,
$\pi_0\equiv\pi_{\rm NS}^+$ ($\pi_0$ being the
basic, i.e.~vacuum representation) and
$\pi_{\rm v}\equiv\pi_{\rm NS}^-$. Consider
the Bogoliubov operator
$V_{1/2}\in{\cal I}({\cal K},\Gamma)$,
\[ V_{1/2} = \sum_{i=1}^N \left(
\frac{1}{\sqrt{2}} |e_\frac{1}{2}^i\rangle\langle
e_0^i| + \frac{1}{\sqrt{2}} |e_{-\frac{1}{2}}^i
\rangle\langle e_0^i| + \sum_{n=1}^\infty
\big( |e_{n+\frac{1}{2}}^i\rangle\langle e_n^i|
+ |e_{-n-\frac{1}{2}}^i\rangle\langle e_{-n}^i|
\big) \right). \]
It is not hard to see that
$S_{\rm R}=V_{1/2}^*P_{\rm NS}V_{1/2}$, that
$M_{V_{1/2}}=N$ and that $N_{V_{1/2}}=0$. We
find
by Eq.~(\ref{cyclics})
$\pi_{\rm R}\simeq\pi_{\rm NS}\circ\varrho_{V_{1/2}}$
by Eq.~(\ref{cyclics}), and hence by Theorem
\ref{restodd},
\begin{equation}
\label{R+}
\pi_{\rm R}|_{{\cal C}({\cal K},\Gamma)^+}
\simeq \left\{ \begin{array}{cl}
2^\frac{N}{2}(\pi_{P'}^+\oplus\pi_{P'}^-) &
\qquad N\in 2{\mathbb N} \\
2^\frac{N+1}{2} \pi & \qquad N\in 2{\mathbb N}_0+1
\end{array} \right.
\end{equation}
for a basis projection $P'$,
$[P']_2=[S_{\rm R}^\frac{1}{2}]_2$. Thus we use
notations $\pi_{\rm s}\equiv\pi_{P'}^+$,
$\pi_{\rm c}\equiv\pi_{P'}^-$ and
$\pi_\sigma\equiv\pi$. (Recall that $\pi$ is one of
the equivalent restrictions of the pseudo Fock
representations $\pi_{E,\pm}$.) We have seen that the
CAR representations $\pi_{\rm NS}$ and $\pi_{\rm R}$,
when restricted to the even algebra, reproduce
precisely the sectors of the chiral algebra. This
is not a surprise because the Kac-Moody and Virasoro
generators are made of fermion bilinears. Note that the
Bogoliubov endomorphism $\varrho_{V_{1/2}}$
induces a transition from the vacuum sector to
spinor sectors.

Let us finish this section with some brief remarks
on M{\"o}bius covariance of the
vacuum sector. The M{\"o}bius symmetry on the circle
$S^1$ is given by the group
${\rm PSU}(1,1)={\rm SU}(1,1)/{\mathbb Z}_2$ where
\[ {\rm SU}(1,1) = \left\{ \left. g= \left(
\begin{array}{cc}
\alpha & \beta \\ \overline{\beta} & \overline{\alpha}
\end{array} \right) \in {\rm GL}_2({\mathbb C}) \,\,
\right| \,\, |\alpha|^2-|\beta|^2=1 \right\}. \]
Its action on the circle is
\[ gz= \frac{\overline{\alpha}z-
\overline{\beta}}{-\beta z + \alpha}, \qquad
z \in S^1. \]
Consider the one-parameter-group of rotations
$a_0(t)$,
\[ a_0(t) = \left( \begin{array}{cc}
\E^{-\frac{\I t}{2}} & 0 \\ 0 &
\E^\frac{\I t}{2} \end{array} \right),
\qquad t \in {\mathbb R}. \]
Any element $g\in{\rm SU}(1,1)$ can be decomposed
in a rotation $a_0(t)$ and a transformation
$g'=a_0(-t)g$ leaving the point $z=-1$ invariant,
\[ g=a_0(t)g', \qquad  g'= \left(
\begin{array}{cc} \alpha' & \beta' \\
\overline{\beta'} & \overline{\alpha'}
\end{array} \right), \qquad
\frac{\overline{\alpha'}+
\overline{\beta'}}{\alpha' + \beta'}=1. \]
Since $a_0(t+2\pi)=-a_0(t)$ we can determine
$t$, $-2\pi<t\le 2\pi$ uniquely by the
additional requirement ${\rm Re}(\alpha')>0$. Then
a representation $U$ of ${\rm SU}(1,1)$ in our
Hilbert space ${\cal K}$ of test functions
$f=(f^i)_{i=1,\ldots,N}$ is defined component-wise by
\[ \left( U (g) f \right)^i (z) = \epsilon(g;z)
(\alpha + \overline{\beta}
\overline{z})^{-\frac{1}{2}}(\overline{\alpha}
+\beta z)^{-\frac{1}{2}}
f^i \left( \frac{\alpha z + \overline{\beta}}{\beta z
+ \overline{\alpha}} \right)  \]
where for $z=\E^{\I\phi}$, $-\pi<\phi\le\pi$
\[ \epsilon(g;z)= -\,{\rm sign}(t-\pi-\phi)
\,\,{\rm sign}(t+\pi-\phi), \]
and ${\rm sign}(x)=1$ if $x\ge 0$,
${\rm sign}(x)=-1$ if $x<0$. By the same
arguments as in the appendix
of \cite{leci} for the case $N=1$ one checks
that $U$ is indeed a well-defined and unitary
representation. Moreover, since the prefactor
on the right hand side
is real we observe $[U(g),\Gamma]=0$ and hence
each $U(g)$, $g\in{\rm SU}(1,1)$, induces a
Bogoliubov automorphism
$\alpha_g=\varrho_{U(g)}$.
It follows that ${\rm SU}(1,1)$ is represented
by automorphisms of ${\cal C}({\cal K},\Gamma)$,
and this restricts to a representation of
${\rm PSU}(1,1)$ by automorphisms of
${\cal C}({\cal K},\Gamma)^+$. Again, as an obvious
generalization of the computations in \cite{leci}
one checks $[P_{\rm NS},U(g)]=0$ and hence
$\pi_{\rm NS}\circ\alpha_g\simeq\pi_{\rm NS}$
and $\pi_0\circ\alpha_g\simeq\pi_0$.
Consider further one-parameter-subgroups
\[ a_+(t) = \left( \begin{array}{cc}
\cosh t & \I \, \sinh t \\ - \I \, \sinh t &
\cosh t \end{array} \right), \qquad
a_-(t) = \left( \begin{array}{cc}
\cosh t &  \sinh t \\ - \sinh t &
\cosh t \end{array} \right), \]
$t\in{\mathbb R}$.
It is not hard to check that the $a_\epsilon$
correspond to infinitesimal generators
$d_0$, $d_+=d_1+d_{-1}$
and $d_-=-\I(d_1-d_{-1})$ by
$U(a_\epsilon(t))=\exp (\I t d_\epsilon)$,
$\epsilon=0,\pm$, where
\[ d_n = - \sum_{i=1}^N \sum_{r\in{\mathbb Z}
+\frac{1}{2}} \left( r+\frac{n}{2} \right)
|e_{r+n}^i \rangle\langle e_r^i |, \qquad
n= 0, \pm 1, \]
i.e. the $d_n$ act as $-z^n \left( z
\frac{{\rm d}}{{\rm d}z} + \frac{n}{2} \right)$
in each component. The relation
\[ [L_n^{\rm NS},b_r^i]= - \left( r+\frac{n}{2}
\right) b_{r+n}^i, \qquad r\in{\mathbb Z}
+\frac{1}{2}, \quad i=1,2,\ldots,N, \quad
n=0, \pm 1, \]
establishes the correspondence between generators
$d_\epsilon$ in the test function space ${\cal K}$
and infinitesimal generators $L_\epsilon^{\rm NS}$
in the Fock space ${\cal H}_{\rm NS}$; we have
(for $f$ in the domain of $d_\epsilon$)
\[ [L_\epsilon^{\rm NS}, \pi_{\rm NS}(B(f))]
= \pi_{\rm NS}(B(d_\epsilon f)), \qquad
\epsilon = 0, \pm, \]
where $L_+^{\rm NS}=L_1^{\rm NS}+L_{-1}^{\rm NS}$
and $L_-^{\rm NS}=-\I(L_1^{\rm NS}-L_{-1}^{\rm NS})$.

\section{Localized Endomorphisms}
In this section we construct localized endomorphisms
by means of Bogoliubov endomorphisms. Thus we have to
introduce at first a local structure on $S^1$,
i.e.~to define local algebras of observables.
Let us denote by ${\cal J}$ the set of open,
non-void proper subintervals of $S^1$. For
$I\in{\cal J}$ set
${\cal K}(I)=L^2(I;{\mathbb C}^N)$
and define local $C^*$-algebras
\[ {\cal A}(I)={\cal C}({\cal K}(I),\Gamma)^+ \]
such that we have inclusions
\[ {\cal A}(I) \subset {\cal A}(I_0), \qquad
I \subset I_0, \]
inherited by the natural embedding of the
$L^2$-spaces; and also we have locality,
\[ [{\cal A}(I),{\cal A}(I_1)]=\{0\},
\qquad I\cap I_1 = \emptyset. \]
Our construction of localized endomorphisms
happens on the punctured circle. Consider
the interval $I_\zeta\in{\cal J}$ which is $S^1$
by removing one ``point at infinity''
$\zeta\in S^1$, $I_\zeta=S^1\setminus\{\zeta\}$.
Clearly,
${\cal A}(I_\zeta)={\cal C}({\cal K},\Gamma)^+$.
Further denote by ${\cal J}_\zeta$ the set of
``finite'' intervals $I\in{\cal J}$ such that
their closure is contained in $I_\zeta$,
\[ {\cal J}_\zeta = \{ I\in{\cal J} \,\,|\,\,
\bar{I}\subset I_\zeta \}. \]
An endomorphism $\varrho$ of ${\cal A}(I_\zeta)$
is called localized in some interval
$I\in{\cal J}_\zeta$ if it satisfies
\[ \varrho(A)=A, \qquad A\in{\cal A}(I_1), \qquad
I_1\in{\cal J}_\zeta, \qquad I_1 \cap I = \emptyset. \]
The construction of localized endomorphisms by
means of Bogoliubov transformations leads to
the concept of pseudo-localized isometries
\cite{MS1}. For $I\in{\cal J}_\zeta$ denote
by $I_+$ and $I_-$ the two connected components
of $I'\cap I_\zeta$ ($I'$ always denotes the
interior of the complement of $I$ in $S^1$,
$I'=I^{\rm c}\setminus\partial I^{\rm c}$).
A Bogoliubov operator
$V\in{\cal I}({\cal K},\Gamma)$ is called even
(resp.~odd) pseudo-localized in $I\in{\cal J}_\zeta$
if
\[ Vf = \epsilon_\pm f,\qquad  f\in {\cal K}(I_\pm),
\qquad \epsilon_\pm \in \{-1,1\}, \]
and $\epsilon_+=\epsilon_-$
(resp.~$\epsilon_+=-\epsilon_-$).
Then, as obvious, $\varrho_V$ is localized in $I$
in restriction to ${\cal A}(I_\zeta)$.
Now we are ready to define our localized vector
endomorphism.
\begin{definition}
\label{rhov}
For some $I\in{\cal J}_\zeta$ choose a real
$v\in{\cal K}(I)$, $\Gamma v=v$ and $\|v\|=1$.
Define the unitary self-adjoint Bogoliubov
operator $U\in{\cal I}({\cal K},\Gamma)$ by
\begin{equation}
U= 2|v\rangle\langle v|-{\bf 1},
\end{equation}
and the localized vector endomorphism
(automorphism) $\varrho_{\rm v}$ by
$\varrho_{\rm v}=\varrho_U$.
\end{definition}
Since $U$ is even pseudo-localized, and by
Corollary \ref{Bogvec}, $\varrho_{\rm v}$ is
indeed a localized vector endomorphism,
i.e.~$\pi_0\circ\varrho_{\rm v}\simeq
\pi_{\rm v}$.
Further, by $U^2={\bf 1}$ we have
$\pi_0\circ\varrho_{\rm v}^2\simeq\pi_0$.
It follows also by Corollary \ref{Bogvec}
that $\pi_{\rm s}\circ\varrho_{\rm v}
\simeq \pi_{\rm c}$.
The construction of a localized spinor
endomorphism is a little bit more costly.
Without loss of generality, we choose
$\zeta=-1$ and the localization region
to be $I_2$,
\[ I_2 = \left\{ z=\E^{\I \phi} \in S^1 \,\,\left|
\,\, -\frac{\pi}{2} <
\phi < \frac{\pi}{2} \right. \right\} \]
such that the connected components $I_\pm$ of
$I_2'\cap I_\zeta$ are given by
\[ I_- = \left\{ z=\E^{\I \phi} \in S^1 \,\,\left| \,\,
-\pi < \phi < -\frac{\pi}{2} \right. \right\}, \quad
I_+ = \left\{ z=\E^{\I \phi} \in S^1 \,\, \left| \,\,
\frac{\pi}{2} < \phi < \pi \right. \right\}.\]
Our Hilbert space ${\cal K}={\cal K}(I_\zeta)$
decomposes into a direct sum,
\[ {\cal K} = {\cal K}(I_-) \oplus {\cal K}(I_2)
\oplus {\cal K}(I_+).\]
By $P_{I_+}$, $P_{I_-}$ we denote the projections onto the
subspaces ${\cal K}(I_+)$, ${\cal K}(I_-)$, respectively.
Define functions on $S^1$ by
\[ f_p(z)= \left\{ \begin{array}{cl}
\sqrt{2}z^{2p} & \qquad z\in I_2 \\
0 & \qquad z\notin I_2 \end{array} \right.,
\qquad p\in \frac{1}{2}{\mathbb Z}, \]
and
\[ f_p^i = f_p \otimes u^i, \qquad
p\in \frac{1}{2} {\mathbb Z}, \qquad i=1,2,\ldots,N,\]
such that we obtain two ONB of the subspace
${\cal K}(I_2)\subset{\cal K}$,
\[ \left\{ f_r^i, r\in {\mathbb Z}+\frac{1}{2},
i=1,2,\ldots,N \right\},\qquad \{ f_n^i, n\in{\mathbb Z},
i=1,2,\ldots,N \}. \]
Now define the odd pseudo-localized Bogoliubov operator
$V\in{\cal I}({\cal K},\Gamma)$,
\begin{eqnarray}
V &=& P_{I_-}-P_{I_+}+ V^{(2)}, \\
V^{(2)} &=& \sum_{j\le N \atop j \,\, {\rm odd}}
(\I r^j + \I R^j) -
\sum_{j\le N \atop j \,\, {\rm even}}
(t^j + \I T^j),\\
r^j &=& \frac{1}{\sqrt{2}} |f_\frac{1}{2}^j
\rangle\langle f_0^j| - \frac{1}{\sqrt{2}}
|f_{-\frac{1}{2}}^j \rangle\langle f_0^j|, \\
R^j &=& \sum_{n=1}^\infty \Big( |f_{n+\frac{1}{2}}^j
\rangle\langle f_n^j| - |f_{-n-\frac{1}{2}}^j
\rangle\langle f_{-n}^j|\Big),\\
t^j &=& \frac{1}{\sqrt{2}} |f_\frac{1}{2}^{j-1}
\rangle\langle f_0^j| + \frac{1}{\sqrt{2}}
|f_{-\frac{1}{2}}^{j-1} \rangle\langle f_0^j|, \\
T^j &=& \sum_{n=1}^\infty\Big( |f_{n-\frac{1}{2}}^j
\rangle\langle f_n^j| - |f_{-n+\frac{1}{2}}^j
\rangle\langle f_{-n}^j|\Big).
\end{eqnarray}
Remark that $V$ is unitary if $N\in 2{\mathbb N}$. In
particular we have
\[ M_V= \left\{ \begin{array}{cl} 0 & \qquad
N\in 2{\mathbb N} \\
1 & \qquad N\in 2{\mathbb N}_0+1 \end{array} \right. .\]
Moreover, we claim
\begin{lemma}
\label{HS}
With notations as above,
\begin{eqnarray}
[(V^*P_{\rm NS}V)^\frac{1}{2}]_2 &=&
[S_{\rm R}^\frac{1}{2}]_2,\\
{[}(V^*V^*P_{\rm NS}VV)^\frac{1}{2}{]}_2 &=&
[P_{\rm NS}]_2.
\end{eqnarray}
\end{lemma}
{\it Proof.} Let us first point out that that we
do not have to take care about the positive square
roots because for any basis projection $P$ and any
Bogoliubov operator $W\in{\cal I}({\cal K},\Gamma)$
with $M_W<\infty$ we have
\[ [ (W^*PW)^\frac{1}{2} ]_2 = [ W^*PW ]_2 \]
since
\begin{eqnarray*}
\| (W^*PW)^\frac{1}{2} - W^*PW \|_2^2 &\le&
\| W^*PW - (W^*PW)^2 \|_1 \\
&=& \| W^*P({\bf 1}-WW^*)PW\|_1 \\
&\le& \|W\|^2 \|P\|^2 \|{\bf 1}-WW^* \|_1 = M_W.
\end{eqnarray*}
Here we used the trace norm and Hilbert Schmidt norm
$\|A\|_n=(\mbox{tr}(A^*A)^\frac{n}{2})^\frac{1}{n}$,
$n=1,2$, respectively, and also an estimate \cite{PS}
\begin{equation}
\label{HST}
\|A^\frac{1}{2}-B^\frac{1}{2}\|_2^2\le\|A-B\|_1,
\qquad A,B \in \mathfrak{B}({\cal K}), \qquad A,B \ge 0.
\end{equation}
It was proven in \cite{leci}, Lemma 3.10, that
\[ V^*P_{\rm NS}V - S_{\rm R}, \quad
VP_{\rm NS}V^* - S_{\rm R}, \quad {V'}^*P_{\rm NS}V'
-S_{\rm R}, \quad V'P_{\rm NS}{V'}^* - S_{\rm R} \]
are Hilbert Schmidt operators for the case
$N=1$, where in our notation
\[ V=P_{I_-} - P_{I_+} + \I r^1 + \I R^1,
\qquad V'=P_{I_-} - P_{I_+} + \I (T^1)^*. \]
For arbitrary $N$ operators
$V^*P_{\rm NS}V-S_{\rm R}$ and
$VP_{\rm NS}V^*-S_{\rm R}$ are just direct sums of
the above Hilbert Schmidt operators (up to finite
dimensional operators), hence we conclude for
arbitrary $N$
\[ V^*P_{\rm NS}V - S_{\rm R} \in
\mathfrak{J}_2({\cal K}), \qquad
VP_{\rm NS}V^* - S_{\rm R} \in
\mathfrak{J}_2({\cal K}). \]
Both relations together imply that
$P_{\rm NS}-V^*V^*P_{\rm NS}VV$ is also Hilbert
Schmidt which proves the lemma, q.e.d.

Hence we conclude
$\pi_{\rm NS}\circ\varrho_V\approx\pi_{\rm R}$.
For $N\in 2{\mathbb N}$ the basis projection
$P'=V^*P_{\rm NS}V$ is as in Eq.~(\ref{R+}). For
$N\in 2{\mathbb N}_0+1$ the representation
$\pi_{\rm NS}\circ\varrho_V$, when restricted
to ${\cal C}({\cal K},\Gamma)^+$, decomposes into
two equivalent irreducibles. With our above
definitions and using Corollary \ref{Bogvec},
this suggests the following
\begin{definition}
\label{rhos}
Choose $U\in{\cal I}({\cal K},\Gamma)$ for
$v\in{\cal K}(I_2)$ as in Definition
\ref{rhov}. For $N\in 2{\mathbb N}$ define the
localized spinor endomorphism $\varrho_{\rm s}$
by $\varrho_{\rm s}=\varrho_V$ and the
localized conjugate spinor endomorphism
$\varrho_{\rm c}$ by
$\varrho_{\rm c}=\varrho_U\varrho_V$. For
$N\in 2{\mathbb N}_0+1$ define the localized
spinor endomorphism $\varrho_\sigma$ by
$\varrho_\sigma=\varrho_V$.
\end{definition}
Note that this definition fixes the choice,
if $N$ is even, which of the two inequivalent
spinor sectors is called s and which c. There
is no loss of generality because the fusion
rules turn out to be invariant under exchange
of s and c. Indeed, our considerations have shown
\begin{theorem}
The localized endomorphisms of Definitions
\ref{rhov} and \ref{rhos} satisfy
$\pi_0\circ\varrho_J\simeq\pi_J$,
$J={\rm v,s,c},\sigma$.
\end{theorem}

\section{Extension to Local von Neumann Algebras}
We have obtained the relevant localized endomorphisms
which generate the sectors ${\rm v,s,c},\sigma$. It is
our next aim to derive fusion rules in terms of DHR
sectors i.e.~of unitary equivalence classes
$[\pi_0\circ\varrho]$ for localized endomorphisms
$\varrho$. For such a formulation one needs local
intertwiners in the observable algebra.
So we have to keep close to the DHR framework, in
particular, we should use local von Neumann algebras
instead of local $C^*$-algebras ${\cal A}(I)$.
We define
\[ {\cal R}(I) = \pi_0({\cal A}(I))'',\qquad
I\in{\cal J}. \]
By M\"obius covariance of the vacuum state, this
defines a so-called covariant precosheaf on the
circle \cite{BGL}. In particular, we have Haag duality,
\begin{equation}
\label{HD}
{\cal R}(I)'={\cal R}(I').
\end{equation}
Since the set ${\cal J}$ is not directed by inclusion we
cannot define a global algebra as the $C^*$-norm
closure of the union of all local algebras. However,
the set ${\cal J}_\zeta$ is directed so that we can define
the following algebra $\mathfrak{A}_{\rm loc}$ of
quasilocal observables in the usual manner,
\begin{equation}
\label{Aloc}
\mathfrak{A}_{\rm loc} =
\overline{\bigcup_{I\in{\cal J}_\zeta} {\cal R}(I)}.
\end{equation}
We want to prove that Haag duality holds also on the
punctured circle and need some technical preparation.
Recall that a function $k\in L^2(S^1)$ is in the Hardy
space $H^2$ if $\langle e_{-n},k \rangle =0$ for all
$n\in {\mathbb N}$ where $e_{-n}(z)=z^{-n}$. There is
a Theorem of Riesz (\cite{Doug}, Th.~6.13) which states
that $k(z)\neq 0$ almost everywhere if $k\in H^2$ is
non-zero. Now suppose $f\in P_{\rm NS}{\cal K}$. Then
$g^i\in H^2$ where $g^i(z)=z^\frac{1}{2}
\overline{f^i(z)}$ component-wise, $i=1,2,\ldots,N$.
We conclude
\begin{lemma}
\label{Hardy}
If $f\in P_{\rm NS}{\cal K}$ then $f\in{\cal K}(I)$
implies $f=0$ for any $I\in{\cal J}$.
\end{lemma}
For some interval $I\in{\cal J}_\zeta$, let us denote by
$\mathfrak{A}_\zeta(I')$ the norm closure of the algebra
generated by all
${\cal R}(I_1)$, $I_1\in{\cal J}_\zeta$,
$I_1\cap I=\emptyset$. Obviously
$\mathfrak{A}_\zeta (I')''\subset {\cal R}(I')$; a key
point of the analysis is the following
\begin{lemma}
Haag duality remains valid on the punctured circle, i.e.
\begin{equation}
\label{HDpunc}
{\cal R}(I)'=\mathfrak{A}_\zeta(I')''.
\end{equation}
\end{lemma}
{\it Proof.} We have to prove
$\mathfrak{A}_\zeta(I')''={\cal R}(I')$. It is
sufficient to show that each generator
$\pi_0(B(f)B(g))$, $f,g\in{\cal K}(I')$ of
${\cal R}(I')$ is a weak limit point of a net
in $\mathfrak{A}_\zeta(I')$. Note that the subspace
${\cal K}^{(\zeta)}(I')\subset{\cal K}(I')$ of
functions which vanish in a neighborhood of
$\zeta$ is dense. So by Eq. (\ref{Cnorm}) we
conclude that it is sufficient to establish
this fact only for such generators with
$f,g\in{\cal K}^{(\zeta)}(I')$, because these
generators approximate the arbitrary ones already
in the norm topology. Let us again denote the
two connected components of $I'\setminus\{\zeta\}$
by $I_+$ and $I_-$, and the projections onto
corresponding subspaces ${\cal K}(I_\pm)$ by
$P_\pm$. We also write $f_\pm=P_\pm f$ and
$g_\pm = P_\pm g$ for our functions
$f,g\in{\cal K}^{(\zeta)}(I')$.
Then we have
\begin{eqnarray*}
\pi_0(B(f)B(g)) &=& \pi_0(B(f_+)B(g_+)) +
\pi_0(B(f_-)B(g_-)) \\
&&  \qquad + \pi_0(B(f_+)B(g_-)) +
\pi_0(B(f_-)B(g_+)).
\end{eqnarray*}
Clearly, the first two terms on the r.h.s. are
elements of $\mathfrak{A}_\zeta(I')$. We show that
the third term $Y=\pi_0(B(f_+)B(g_-))$ (then, by
symmetry, also the fourth one) is in
$\mathfrak{A}_\zeta(I')''$. In the same way as in
the proof of Lemma 4.1 in \cite{leci} one
constructs a sequence $\{X_n,n\in{\mathbb N}\}$,
\[ X_n = \pi_0 (B(h_n^+)B(h_n^-)) \]
where unit vectors $h_n^\pm\in{\cal K}(I_n^\pm)$
are related by M{\"o}bius transformations such that
intervals $I_n^\pm\subset I_\pm$ shrink to the point
$\zeta$. Since $\|X_n\|\le 1$ by Eq.~(\ref{Cnorm}) it
follows that there is a weakly convergent subnet
$\{Z_\alpha,\alpha\in\iota\}$ ($\iota$ a directed
set), ${\rm w-}\lim_\alpha Z_\alpha=Z$. For each
$I_0\in{\cal J}_\zeta$ elements $X_n$ commute with
each $A\in{\cal R}(I_0)$ for sufficiently large
$n$. Hence $Z$ is in the commutant of
$\mathfrak{A}_{\rm loc}$ and this implies
$Z=\lambda {\bf 1}$. We have chosen the vectors
$h_n^\pm$ related by M{\"o}bius transformations.
By M{\"o}bius invariance of the vacuum state
we have
\[ \lambda = \langle \Omega_0 | X_1 | \Omega_0
\rangle = \langle \Gamma h_1^+, P_{\rm NS}
h_1^- \rangle . \]
We claim that we can choose $h_1^\pm$ such
that $\lambda\neq 0$. For given $h_1^-$ set
$k=P_{\rm NS}h_1^-$. We have $k\neq 0$, otherwise
$\Gamma h_1^-\in P_{\rm NS}{\cal K}$ in contradiction
to $h_1^-\in{\cal K}(I_1^-)$ by Lemma \ref{Hardy}.
Again by Lemma \ref{Hardy} we conclude that $k$
cannot vanish almost everywhere. So we clearly
can choose a $h_1^+\in{\cal K}(I_1^+)$ such that
$\lambda=\langle\Gamma h_1^+,k\rangle\neq 0$.
Now we find
$Y=\lambda^{-1}{\rm w-}\lim_\alpha YZ_\alpha$
and also $YZ_\alpha\in\mathfrak{A}_\zeta(I')$
because
\[ YX_n = \pi_0(B(f_+)B(g_-)B(h_n^+)B(h_n^-))
= -\pi_0(B(f_+)B(h_n^+)) \pi_0(B(g_-)B(h_n^-)) \]
is in $\mathfrak{A}_\zeta(I')$ for all
$n\in{\mathbb N}$, q.e.d.

Since the vacuum representation is faithful
on ${\cal A}(I_\zeta)$ we can
identify observables $A$ in the usual manner with
their vacuum representers $\pi_0(A)$. Thus we
consider the vacuum representation as acting
as the identity on $\mathfrak{A}_{\rm loc}$, and, in
the same fashion, we treat local $C^*$-algebras
as subalgebras ${\cal A}(I)\subset{\cal R}(I)$.
Now we have to check whether we can extend our
representations $\pi_J$ and endomorphisms
$\varrho_J$ from ${\cal A}(I)$ to
${\cal R}(I)={\cal A}(I)''$, $I\in{\cal J}_\zeta$,
$J={\rm v,s,c},\sigma$. That is that we have to
check local quasiequivalence of the
representations $\pi_J$ and will now be elaborated.
Define $E_{\rm R}\in\mathfrak{B}({\cal K})$ by
\[ E_{\rm R} = \sum_{i=1}^N \sum_{n\in{\mathbb N}}
|e_{-n}^i \rangle\langle e_{-n}^i| +
\sum_{j \le N \atop j \,\, {\rm even}}
|e_+^j\rangle\langle e_+^j| \]
where $e_+^j=2^{-\frac{1}{2}}(e_0^j+\I e_0^{j-1})$.
\begin{lemma}
\label{dense}
For $I\in{\cal J}$ the subspaces
$P_{\rm NS}{\cal K}(I)\subset P_{\rm NS}{\cal K}$
and $E_{\rm R}{\cal K}(I)\subset E_{\rm R}{\cal K}$
are dense.
\end{lemma}
{\it Proof.} Suppose that $P_{\rm NS}{\cal K}(I)$
is not dense in $P_{\rm NS}{\cal K}$. Then there
is a non-zero $f\in P_{\rm NS}{\cal K}$ such that
\[ \langle f, P_{\rm NS} g \rangle =
\langle f, g \rangle = 0 \]
for all $g\in {\cal K}(I)$. Hence
$f\in{\cal K}(I)^\perp ={\cal K}(I')$ in
contradiction to Lemma \ref{Hardy}. As quite
obvious, Lemma \ref{Hardy} holds for
$f\in E_{\rm R}{\cal K}$ as well. So also
$E_{\rm R}{\cal K}(I)$ is dense in
$E_{\rm R}{\cal K}$, q.e.d.

Note that $E_{\rm R}$ is a basis projection if
$N$ is even. For $N$ odd, $E_{\rm R}$ is a partial
basis projection with $\Gamma$-codimension 1 and
corresponding $\Gamma$-invariant unit vector $e_0^N$.
In this case
\[ S_{\rm R}' = \frac{1}{2} |e_0^N \rangle
\langle e_0^N| + E_{\rm R} \]
is of the form (\ref{mittel}). Let us denote by
$({\cal H}_{\rm R'},\pi_{\rm R'},
|\Omega_{\rm R'}\rangle)$ the GNS representation of
the quasifree state $\omega_{E_{\rm R}}$ if $N$ is
even and $\omega_{S_{\rm R}'}$ if $N$ is odd.
We conclude
\[ \pi_{\rm R'}|_{{\cal C}({\cal K},\Gamma)^+}
\simeq \left\{ \begin{array}{cl}
\pi_{\rm s} \oplus \pi_{\rm c} & N\in 2 {\mathbb N}\\
2 \pi_\sigma & N\in 2 {\mathbb N}_0+1 \end{array}
\right. \]
by Theorem \ref{resteven} and Lemma \ref{pFock} and
the fact that $[ E_{\rm R} ]_2 = [ S_{\rm R}^\frac{1}{2}
]_2 = [ S_{\rm R} ]_2$ ($N$ even) and
$[ {S_{\rm R}'}^\frac{1}{2} ]_2 = [ S_{\rm R}' ]_2
= [ S_{\rm R} ]_2$ ($N$ odd).
\begin{lemma}
\label{locquasi}
For $I\in{\cal J}_\zeta$ we have local quasiequivalence
\begin{equation}
\pi_{\rm NS}|_{{\cal C}({\cal K}(I),\Gamma)} \approx
\pi_{\rm R'}|_{{\cal C}({\cal K}(I),\Gamma)}.
\end{equation}
\end{lemma}
{\it Proof.} We first claim that
$|\Omega_{\rm NS}\rangle$ and $|\Omega_{\rm R'}\rangle$
remain cyclic for
$\pi_{\rm NS}({\cal C}({\cal K}(I),\Gamma))$ and
$\pi_{\rm R'}({\cal C}({\cal K}(I),\Gamma))$,
respectively. By Lemma \ref{dense},
$P_{\rm NS}{\cal K}(I)\subset P_{\rm NS}{\cal K}$
is dense. Hence vectors
$\pi_{\rm NS}(B(f_1)\cdots B(f_n))
|\Omega_{\rm NS}\rangle$, with
$f_1,\dots,f_n\in P_{\rm NS}{\cal K}(I)$,
$n=0,1,2,\ldots,$
are total in ${\cal H}_{\rm NS}$. This proves
the required cyclicity of $|\Omega_{\rm NS}\rangle$.
For $N$ even, cyclicity of $|\Omega_{\rm R'}\rangle$
for ${\cal C}({\cal K}(I),\Gamma)$ is proven in
the same way. For $N$ odd, we have
${\cal H}_{\rm R'}={\cal H}_{E_{\rm R}} \oplus
{\cal H}_{E_{\rm R}}$, $\pi_{\rm R'} =\pi_{E,+}
\oplus \pi_{E,-}$ and $|\Omega_{\rm R'} \rangle =
2^{-\frac{1}{2}} (|\Omega_{E_{\rm R}} \rangle
\oplus |\Omega_{E_{\rm R}} \rangle)$ as in Lemma
\ref{pFock}, and the corresponding
$\Gamma$-invariant unit vector is given by
$e_0^N$. In order to prove cyclicity of
$|\Omega_{\rm R'} \rangle$ we show that
$\langle \Psi | \pi_{\rm R'}(x) |
\Omega_{\rm R'} \rangle =0$ for all
$x\in{\cal C}({\cal K}(I),\Gamma)$,
$|\Psi\rangle=|\Psi_+\rangle\oplus|\Psi_-\rangle
\in{\cal H}_{\rm R'}$, implies $|\Psi\rangle=0$.
We have
\[ \langle \Psi | \pi_{\rm R'}(x) |
\Omega_{\rm R'} \rangle = \frac{1}{\sqrt{2}}
\langle \Psi_+ | \pi_{E_{\rm R},+}(x) |
\Omega_{E_{\rm R}} \rangle + \frac{1}{\sqrt{2}}
\langle \Psi_- | \pi_{E_{\rm R},-}(x) |
\Omega_{E_{\rm R}} \rangle = 0 \]
Again by Lemma \ref{dense},
$E_{\rm R}{\cal K}(I)\subset E_{\rm R}{\cal K}$
is dense, hence vectors
$\pi_{E_{\rm R},\pm}(x)|\Omega_{E_{\rm R}}\rangle =
\pi_{E_{\rm R}}(x)|\Omega_{E_{\rm R}}\rangle$,
$x=B(f_1)\cdots B(f_n)$,
$f_1,\ldots,f_n\in E_{\rm R}{\cal K}(I)$,
$n=0,1,2,\ldots,$ are total in ${\cal H}_{E_{\rm R}}$.
It follows $|\Psi_-\rangle=-|\Psi_+\rangle$.
Hence
\[ \langle \Psi_+ | (\pi_{E_{\rm R},+}(y)-
\pi_{E_{\rm R},-}(y))| \Omega_{E_{\rm R}} \rangle
= 0, \qquad y\in{\cal C}({\cal K}(I),\Gamma). \]
Keep all $x=B(f_1)\cdots B(f_n)$ as above and
choose an $f\in{\cal K}(I)$ such that
$\langle e_0^N,f\rangle=2^{-\frac{1}{2}}$.
Set $y=(-1)^n B(f)x$. Then, by Eq.~(\ref{piE}),
we compute
\[ \pi_{E_{\rm R},\pm}(y)= (-1)^n \left(
\pm \frac{1}{2} Q_{E_{\rm R}}(-1) +
\pi_{E_{\rm R}}(B((E_{\rm R}+
{\overline{E}}_{\rm R})f)) \right)
\pi_{E_{\rm R}}(x) \]
and hence
\[ (\pi_{E_{\rm R},+}(y)- \pi_{E_{\rm R},-}(y))
| \Omega_{E_{\rm R}} \rangle =
\pi_{E_{\rm R}}(B(f_1)\cdots B(f_n)| \Omega_{E_{\rm R}}
\rangle . \]
Because such vectors are total in
${\cal H}_{E_{\rm R}}$ we find $|\Psi_+\rangle=0$
and hence $|\Psi\rangle=0$. We
have seen that vectors $|\Omega_{\rm NS}\rangle$
and $|\Omega_{\rm R'}\rangle$ remain cyclic.
Thus we can prove the lemma by showing
that the restricted states
$\omega_{P_IP_{\rm NS}P_I}$ and
$\omega_{P_IE_{\rm R}P_I}$ ($N\in 2{\mathbb N}$)
respectively $\omega_{P_IS_{\rm R}'P_I}$
($N\in 2{\mathbb N}_0+1$) give rise to
quasiequivalent representations. Because
they are quasifree on ${\cal C}({\cal K}(I),\Gamma)$
we have to show that
\[ [ (P_IP_{\rm NS}P_I)^\frac{1}{2} ]_2 = \left\{
\begin{array}{cl} {[} (P_IE_{\rm R}P_I)^\frac{1}{2}
{]_2} & \qquad N\in 2 {\mathbb N} \\
{[} (P_IS_{\rm R}'P_I)^\frac{1}{2} {]_2} & \qquad
N\in 2 {\mathbb N}_0+1 \end{array} \right..\]
By use of Eq.~(\ref{HST}) it is sufficient to
show that the difference of $P_IP_{\rm NS}P_I$
and $P_IE_{\rm R}P_I$ respectively
$P_IS_{\rm R}'P_I$ is trace class. It was
proven in \cite{leci} that for
$I\in{\cal J}_\zeta$ the difference
$P_IP_{\rm NS}P_I-P_IS_{\rm R}P_I$ is trace
class in the case $N=1$. The result follows
because the operators above are, up to
finite dimensional operators, direct sums
of those for $N=1$, q.e.d.

In restriction to the local even algebra
${\cal A}(I)={\cal C}({\cal K}(I),\Gamma)^+$,
$I\in{\cal J}_\zeta$ we find by Lemma
\ref{locquasi}
\[ (\pi_0\oplus\pi_{\rm v})|_{{\cal A}(I)}
\approx \left\{ \begin{array}{cl}
(\pi_{\rm s} \oplus \pi_{\rm c})
|_{{\cal A}(I)} & \qquad N\in 2 {\mathbb N}\\
2 \pi_\sigma |_{{\cal A}(I)} & \qquad
N\in 2 {\mathbb N}_0+1 \end{array}
\right. \]
Recall that $\pi_{\rm v}\simeq\pi_0\circ\varrho_U$
with $U=2|v\rangle\langle v|-{\bf 1}$ as in
Corollary \ref{Bogvec}. Choose $v\in{\cal K}(I')$.
Then $\varrho_U(x)=x$ for $x\in{\cal A}(I)$, hence
$\pi_0$ and $\pi_{\rm v}$ are equivalent on
${\cal A}(I)$. In the same way we obtain local
equivalence of $\pi_{\rm s}$ and $\pi_{\rm c}$.
We conclude
\begin{theorem}[Local Normality]
\label{locnorm}
In restriction to local $C^*$-algebras ${\cal A}(I)$,
$I\in{\cal J}_\zeta$, the representations $\pi_J$ are
quasiequivalent to the vacuum representation
$\pi_0=id$,
\begin{equation}
\pi_J|_{{\cal A}(I)} \approx \pi_0|_{{\cal A}(I)},
\qquad I\in{\cal J}_\zeta,\quad J={\rm v,s,c},\sigma.
\end{equation}
\end{theorem}
We have seen that we have an extension of our
representations $\pi_J$ to local von Neumann
algebras ${\cal R}(I)$, $I\in{\cal J}_\zeta$, and
thus to the quasilocal algebra
$\mathfrak{A}_{\rm loc}$ they generate. By unitary
equivalence $\varrho_J\simeq\pi_J$ on
${\cal A}(I_\zeta)$ we have an extension of
$\varrho_J$ to $\mathfrak{A}_{\rm loc}$, too,
$J={\rm v,s,c},\sigma$. Being localized in
some $I\in{\cal J}_\zeta$, they inherit
properties
\[ \varrho_J(A)=A, \qquad A\in\mathfrak{A}_\zeta(I'), \]
and
\[ \varrho_J({\cal R}(I_0)) \subset {\cal R}(I_0),
\qquad I_0\in{\cal J}_\zeta, \qquad I\subset I_0, \]
from the underlying $C^*$-algebras. So our
endomorphisms $\varrho_J$, $J={\rm v,s,c},\sigma$
are well-defined localized endomorphisms
$\mathfrak{A}_{\rm loc}$ in the common sense.
Moreover, they are transportable. This follows
because the precosheaf $\{{\cal R}(I)\}$ is
M\"obius covariant. Hence $\mathfrak{A}_{\rm loc}$
is covariant with respect to the subgroup of
M\"obius transformations leaving $\zeta$
invariant.

\section{Fusion Rules}
In this section we prove the fusion rules of our
sectors $1,{\rm v,s,c},\sigma$ in terms of unitary
equivalence classes of localized endomorphisms
$[\varrho]\equiv [\pi_0\circ\varrho]$ (or,
equivalently, in terms of equivalence classes
$[\pi]$ of representations $\pi$ satisfying an
DHR criterion). Because we deal with an Haag dual
net of local von Neumann algebras, by standard
arguments, it suffices to check a fusion rule
$[\varrho_J \varrho_{J'}]$ for special
representatives $\varrho_J\in [\varrho_J]$,
$\varrho_{J'}\in [\varrho_{J'}]$. This will be
done by our examples of Definitions \ref{rhov}
and \ref{rhos}. Let us denote the unitary equivalence
class $[\varrho_J]$ simply by $J$, the fusion
$[\varrho_J\varrho_{J'}]$ by $J\ast J'$ and the
direct sum $[\varrho_J]\oplus [\varrho_{J'}]$ by
$J+J'$, $J,J'=1,{\rm v,s,c},\sigma$. For instance,
we clearly have ${\rm v}\ast{\rm v}=1$ for all
$N\in{\mathbb N}$. Let us first consider the even
case, $N\in 2{\mathbb N}$. By Corollary \ref{Bogvec}
we easily find ${\rm v}\ast{\rm s}={\rm c}$,
${\rm v}\ast{\rm c}={\rm s}$. Since $V$ then is
unitary and by Lemma \ref{HS} we
have $\pi_{\rm NS}\circ\varrho_V^2\simeq\pi_{\rm NS}$.
Now $\pi_{\rm NS}$, when restricted to
${\cal A}(I_\zeta)\equiv{\cal C}({\cal K},\Gamma)^+$,
decomposes into the basic
and the vector representation. Hence only the
possibilities ${\rm s}\ast{\rm s}=1$ or
${\rm s}\ast{\rm s}={\rm v}$ are left, i.e.~we
have to check whether
$\pi_{\rm NS}^+\circ\varrho_V^2$ is equivalent to
$\pi_{\rm NS}^+$ or $\pi_{\rm NS}^-$,
i.e.~whether $\varrho_{\rm s}$ is a
self-conjugate endomorphism or not. For $N$ even
the action of $V$ in the $(2j-1)^\mathrm{th}$ and the
$2j^\mathrm{th}$ component, $j=1,2,\dots,N/2$, is the
same as in the $1^\mathrm{st}$ and the $2^\mathrm{nd}$
component. So we can write the square $W=V^2$ as a
product,
\[ W=W_{1,2}W_{3,4}\cdots W_{N-1,N} \]
where $W_{1,2}$ acts as $W$ in the first two
components and as the identity in the others, etc.
Since $\sigma$ of Prop.~\ref{ind} is multiplicative
and clearly all $W_{2j-1,2j}$ lead to implementable
automorphisms we have
\[ \sigma(W) = \sigma(W_{1,2})\sigma(W_{3,4})
\cdots \sigma(W_{N-1,N}). \]
All $W_{2j-1,2j}$ are built in the same way, hence
all the $\sigma(W_{2j-1,2j})$ are equal
i.e.~$\sigma(W)=\sigma(W_{1,2})^{N/2}$.
Since $\sigma$ takes only values $\pm 1$ this
is ${\rm s}\ast{\rm s}=1$ if $N\in 4{\mathbb N}$.
But for $N\in 4{\mathbb N}_0+2$ this reads
$\sigma(W)=\sigma(W_{1,2})$. Thus we first
check the case $N=2$. If $\sigma(W_{1,2})=+1$
then $\varrho_{\rm s}$ is self-conjugate,
otherwise it is not self-conjugate,
i.e.~${\rm s}\ast{\rm s}={\rm v}$.
It is a result of Guido and Longo \cite{GL} that a
conjugate morphism $\overline{\varrho}$ is given by
\[ \overline{\varrho} =
{\rm j} \circ \varrho \circ {\rm j} \]
where ${\rm j}$ is the antiautomorphism corresponding
to the reflection $z\mapsto\overline{z}$ on the
circle (PCT transformation). In our model,
${\rm j}$ is the extension of the antilinear Bogoliubov
automorphism ${\rm j}_\Theta$,
\[ {\rm j}_\Theta (B(f))=B(\Theta f), \qquad
\Theta f = \Theta \left( (f^i)_{i=1,2}
\right) = \left( \overline{f_{\rm refl}^i}
\right)_{i=1,2}, \]
where $f\in L^2(S^1;{\mathbb C}^2)$
and $\overline{f_{\rm refl}^i}(z)=
\overline{f^i(\overline{z})}$ for $z\in S^1$.
So we have a candidate
$\overline{\varrho_{\rm s}}\equiv\overline{\varrho_V}
=\varrho_{\Theta V\Theta}$. It is quite obvious
that $\Theta P_{I_\pm} \Theta = P_{I_\mp}$ and
that $\Theta f_p^i = f_p^i$, $p\in\frac{1}{2}
{\mathbb Z}$, so it follows by
antilinearity of $\Theta$ ($N=2$)
\[ \Theta V \Theta = -P_{I_-}+P_{I_+} +
(-\I r^1-\I R^1) - (t^2 - \I T^2). \]
It is not hard to see that this is
\[ \Theta V \Theta = U_{1,2} V, \qquad
U_{1,2} = 2 |v_\frac{1}{2}^1\rangle
\langle v_\frac{1}{2}^1|-{\bf 1}, \qquad
v_\frac{1}{2}^1=\frac{1}{\sqrt{2}}
(f_\frac{1}{2}^1+f_{-\frac{1}{2}}^1). \]
Now $U_{1,2}$ is as in Corollary \ref{Bogvec}
so that we find
${\rm s}\ast{\rm v}\ast{\rm s}={\rm s}\ast{\rm c}=1$
for $N=2$. Hence $\sigma(W_{1,2})=-1$, so it follows
${\rm s}\ast{\rm c}=1$ for all $N\in 4{\mathbb N}_0+2$.
For the case $N\in 2{\mathbb N}_0+1$ the situation is
different because $\varrho_V$ then is not an
automorphism. As discussed at the end of Section 5,
the representation $\pi_{\rm NS}\circ\varrho_V$
(and, of course, also
$\pi_{\rm NS}\circ\varrho_U\varrho_V$) decomposes,
in restriction to ${\cal C}({\cal K},\Gamma)^+$,
into two equivalent irreducibles corresponding to
the spinor sector $\sigma$. So we find at first
${\rm v}\ast\sigma=\sigma$. Let us consider
$\pi_{\rm NS}\circ\varrho_V^2$. We have
$M_{V^2}=2M_V=2$, hence by Theorem \ref{evenodd}
and Lemma \ref{HS} we conclude
$\pi_{\rm NS}\circ\varrho_V^2\simeq 2\pi_{\rm NS}$.
In restriction to ${\cal C}({\cal K},\Gamma)^+$
this reads $\pi_{\rm NS}^+\circ\varrho_V^2\oplus
\pi_{\rm NS}^-\circ\varrho_V^2\simeq
2(\pi_{\rm NS}^+\oplus\pi_{\rm NS}^-)$.
Our previous results admit
only $\pi_{\rm NS}^+\circ\varrho_V^2\simeq
\pi_{\rm NS}^-\circ\varrho_V^2$ and hence we find
$\sigma\ast\sigma=1+{\rm v}$. Summarizing we
rediscover the WZW fusion rules.
\begin{theorem}[Fusion Rules]
The basic (1), vector ({\rm v}) and spinor
({\rm s,c},$\sigma$) sectors compose as follows.
Dependent on the integer $N$, the fusion rules read
\begin{eqnarray}
{\rm v}\ast{\rm v}={\rm s}\ast{\rm s}=
{\rm c}\ast{\rm c}=1, \qquad {\rm s}\ast
{\rm c}={\rm v},  \qquad && N\in 4{\mathbb N},\\
{\rm v}\ast{\rm v}={\rm s}\ast{\rm c}=1,
\qquad {\rm s}\ast{\rm s}={\rm c}\ast
{\rm c}={\rm v},  \qquad &&
N\in 4{\mathbb N}_0+2, \\
{\rm v}\ast{\rm v}=1, \qquad \sigma\ast
{\rm v}=\sigma, \qquad \sigma\ast\sigma
=1+{\rm v},  \quad && N\in 2{\mathbb N}_0+1.
\end{eqnarray}
\end{theorem}
We observe that if $N$ is even all sectors are simple.
For $N\in 4{\mathbb N}$ the fusion rules correspond to
the abelian group ${\mathbb Z}_2\times{\mathbb Z}_2$,
for $N\in 4{\mathbb N}_0+2$ they correspond to
${\mathbb Z}_4$. If $N$ is odd the spinor sector
$\sigma$ is not simple corresponding to the fact that
$\varrho_\sigma$ is not an automorphism; one obtains
the Ising fusion rules.

\end{document}